# Transformation of 2-line ferrihydrite to goethite at alkaline pH

Fabio E. Furcas[a], Barbara Lothenbach[b], Shishir Mundra[a], Camelia Borca[c], Cristhiana C. Albert[a], O. Burkan Isgor[d], Thomas Huthwelker[c], Ueli M. Angst[a] *

[a]*Institute for Building Materials, ETH Zürich, Laura-Hezner-Weg 7, 8093, Zürich, Switzerland*

[b]*Empa, Concrete & Asphalt Laboratory, Switzerland*

[c]*Swiss Light Source, Paul Scherrer Institut, 5232 Villigen, Switzerland*

[d]*School of Civil and Construction Engineering, Oregon State University, Corvallis, OR, USA*

*\*corresponding author: uangst@ethz.ch*

Abstract

The transformation of 2-line ferrihydrite to goethite from supersaturated solutions at alkaline pH ≥ 13.0 was studied using a combination of benchtop and advanced synchrotron techniques such as X-ray diffraction, thermogravimetric analysis and X-ray absorption spectroscopy. In comparison to the transformation rates at acidic to mildly alkaline environments, the half-life, $t_{1/2}$, of 2-line ferrihydrite reduces from several months at pH = 2.0, and approximately 15 days at pH = 10.0, to just under 5 hours at pH = 14.0. Calculated first order rate constants of transformation, k, increase exponentially with respect to the pH and follow the progression $\log_{10} k = \log_{10} k_0 + a \cdot pH^3$. Simultaneous monitoring of the aqueous Fe(III) concentration via inductively coupled plasma optical emission spectroscopy demonstrates that (i) goethite likely precipitates from solution and (ii) its formation is rate-limited by the comparatively slow re-dissolution of 2-line ferrihydrite. The analysis presented can be used to estimate the transformation rate of naturally occurring 2-line ferrihydrite in aqueous electrolytes characteristic to mine and radioactive waste tailings as well as the formation of corrosion products in cementitious pore solutions.

*Keywords: Precipitation, iron, kinetics, pH, XAS, XRD, ICP, TGA*

Graphical Abstract

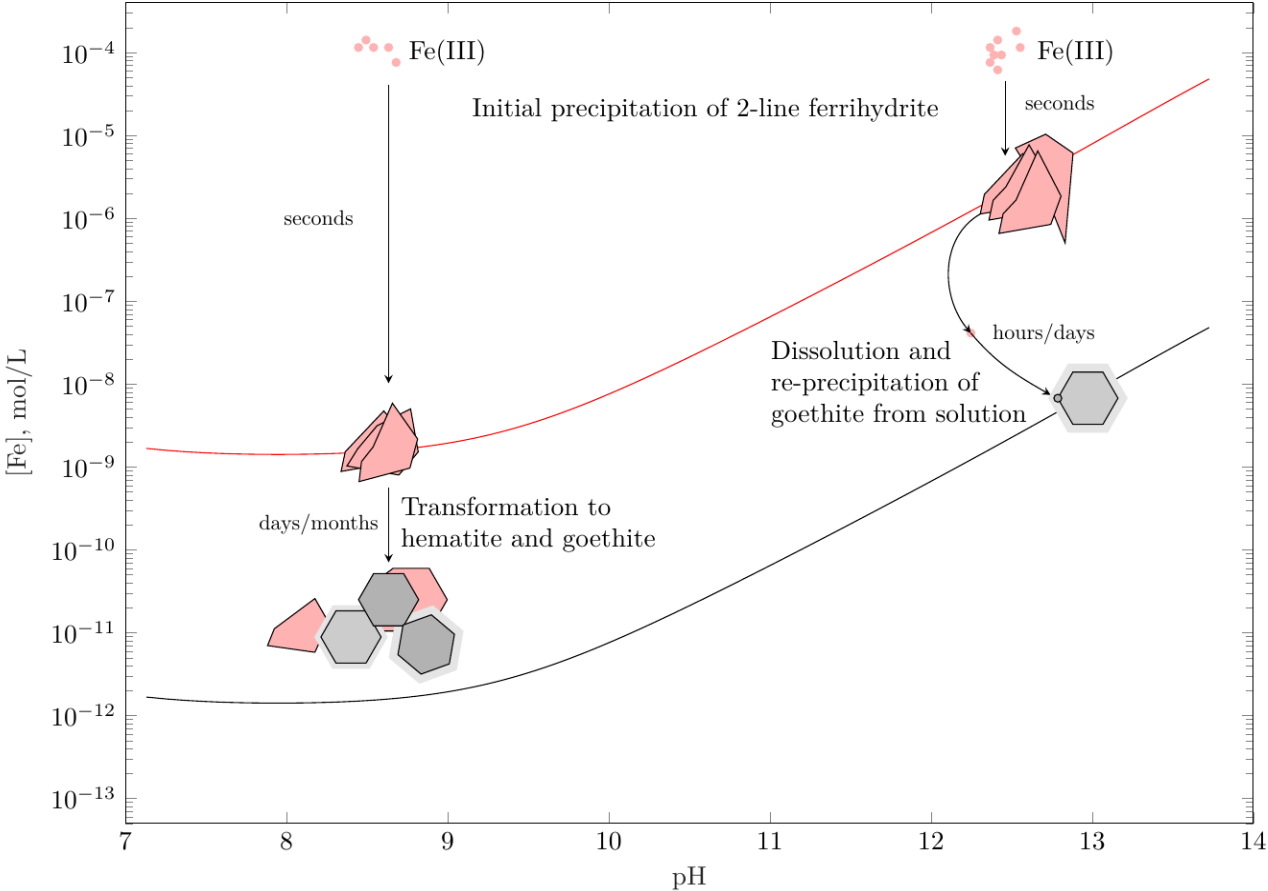

# 1. Introduction

Amorphous Fe(III) (hydr)oxide intermediates and end members can be ranked according to their thermodynamic stability. 2-line ferrihydrite (2l-Fe(OH)$_3$(s)), a nanocrystalline iron hydroxide, is generally recognised as the least stable naturally occurring iron-bearing phase. The absence of a long-range order of arrangement renders it thermodynamically unstable compared to other, more crystalline iron (hydr)oxides. Due to its remarkable sorption capacity and high surface area [1-3], 2-line ferrihydrite readily absorbs groundwater contaminants [4-6] and is commonly used in a range of industrial applications including heavy metal sequestration and nitrogen removal from wastewater [7-10]. Its formation and transformation to other iron (hydr)oxides dictate the amount of iron that remains mobile in the aqueous phase, prospectively impacting the long-term performance of nuclear waste repositories and reinforced concrete structures [11-16]. Since most of the highly crystalline ferric (hydr)oxide phases can coexist with or be synthesized from 2-line ferrihydrite [2, 17-19], it can be considered a gateway molecule that plays an important role in the kinetic mechanism leading to their formation.

Phase transformation of 2-line ferrihydrite to hematite and goethite has been investigated extensively at low (pH = 2) to moderately alkaline pH (pH = 12) and temperatures ranging from 4 to 100 °C [1, 17-29]. Whilst goethite is preferentially stabilised at low to moderate temperatures and either acidic (pH<6) or alkaline (pH>10) conditions [17, 20, 26], the formation of hematite is favoured at circumneutral pH or elevated temperatures [22, 25, 26, 28]. In the former case, rapid dissolution of 2-line ferrihydrite prompts re-precipitation into goethite from aqueous Fe(OH)$_2^+$ at slightly acidic and Fe(OH)$_4^-$ at alkaline pH [1, 21]. As dissolution of 2-line ferrihydrite is minimal at circumneutral pH, its transformation into hematite proceeds via a two-step crystallisation process with goethite forming as an intermediate [1, 19, 25, 26, 28, 29].

Irrespective of the stabilised end member, transformation rates appear to follow first order kinetics with respect to the amount of 2-line ferrihydrite present [20, 21, 26]. The formation of either hematite or goethite is said to be dissolution controlled. Studies furthermore suggest that the rate of transformation is strictly related to the amount of OH$^-$ present in the system. At acidic pH, this observation is in keeping with the thermodynamic and kinetic aspects constituting the crystallisation process: Firstly, the solubility limit, as dictated by the dissolution of any iron (hydr)oxide species features a minimum at the zero point of charge (ZPC) [20] at ≈ pH 8. At acidic pH, an increase in the OH$^-$ activity towards the ZPC at circumneutral pH thus increases the thermodynamic driving force for primary nucleation of goethite or hematite to occur. Secondly, Fe(III) precipitation proceeds more rapidly if the aqueous OH$^-$ to Fe ratio mimics that of the solid phase stabilised [17, 30, 31]. At alkaline pH, however, the solubility limit of iron is in equilibrium with the aqueous concentration of Fe(OH)$_4^-$, which increases by about one order of magnitude per unit increase in pH[32]. High levels of alkalinity beyond the ZPC thus reduce the initial driving force for precipitation and, once formed, promote the re-dissolution of amorphous iron (hydr)oxide intermediates such as 2-line ferrihydrite.

As the formation of such amorphous or microcrystalline phases generally precedes the stabilisation of their thermodynamically more favourable counterparts [33], the apparent reduction in thermodynamic driving force and the observed increase in the rate of re-dissolution compete with one another. Despite an abundance of literature on the transformation of 2-line ferrihydrite, a detailed investigation into the underlying kinetic mechanism remains challenging for a number of reasons. Firstly, most studies are concerned with the transformation of re-dispersed 2-line ferrihydrite and do not consider the competition between primary nucleation of 2-line ferrihydrite and its re-dissolution [17, 20, 26]. Secondly, solids are often in the presence of highly concentrated impurities that are readily incorporated into amorphous iron-bearing phases, further delaying phase transformation [34, 35]. Lastly, the majority of studies reviewed solely quantify the transformation of 2-line ferrihydrite into hematite/goethite in terms of their solid fractions. As the ability of these iron (hydr)oxides to scavenge As, Sr, Cd and other elements of concern from highly alkaline radioactive waste and air pollution control residues depends on the ionic strength, it is equally important to measure also the simultaneous turnover of aqueous Fe(III) concentration coinciding with the solid phase transformation.

This study investigates the transformation of 2-line ferrihydrite from supersaturation at highly alkaline pH ≥13 via a combination of time-resolved X-ray absorption spectroscopy (XAS), thermogravimetric analysis (TGA), X-ray diffraction (XRD), and inductively coupled plasma optical emission spectroscopy (ICP-OES) studies. The resulting data were used to establish a relationship between the rate of 2-line ferrihydrite transformation and the OH$^-$ activity across a broad range of conditions reaching from the acidic (pH=2) to highly alkaline (pH=14) media. The results obtained will help to model the long-term performance of radioactive waste repositories and give further insights into the formation of corrosion products as well as the stability and sorption capacity of 2-line ferrihydrite in alkaline environments including the pore solution of cementitious systems and radioactive waste tailings.

## 2. Materials and methods

### 2.1. Synthesis of pure iron (hydr)oxide phases

The synthesis of all (hydr)oxide phases used as references follows the recommendations of Schwertmann and Cornell [2]. To enhance product crystallinity above the levels obtained by an Fe(III) iron source and eliminate the system favouring the formation of α-FeOOH(s), γ-FeOOH(s) was synthesized from FeCl$_2$·4H$_2$O(cr) (Sigma-Aldrich, ReagentPlus®, 98 %, CAS: 13478-10-9) at room temperature. 60 mM of FeCl$_2$·4H$_2$O(cr) was added to 300 mL of ultra-pure water (UPW) (18.2 MΩcm) under rigorous stirring. Upon bubbling the starting solution with air at a flow rate of ~ 120 mL min$^{-1}$, lepidocrocite is formed according to

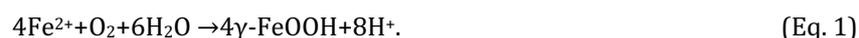

$$4Fe^{2+} + O_2 + 6H_2O \rightarrow 4\gamma\text{-FeOOH} + 8H^+. \qquad \text{(Eq. 1)}$$

To neutralise the protons that formed during the hydrolysis and maintain the solution at a pH of 6.8 ± 0.1, a total of 130 mL 1 M NaOH was added incrementally. The first 100 mL was added drop-wise through a burette, whilst the remainder was pipetted at volumes of 250 µL at a time to enable better pH control. The reaction ran to completion after ~ 2.5 hours, as indicated by a constant pH as well as the characteristic colour change from greenish dark blue to first grey, and then, orange-yellow. The product was centrifuged and dried in an oven at 40 °C for 2 days.

2-line ferrihydrite was synthesized by dissolving 100 mM of Fe(NO$_3$)$_3$· 9H$_2$O(cr) (Sigma-Aldrich, ACS reagent grade, ≥98%, CAS: 7782-61-8) in 500 mL of UPW. Throughout the course of the reaction, 350 mL of 1 M KOH was added to neutralise protons released during Fe$^{3+}$ hydrolysis according to

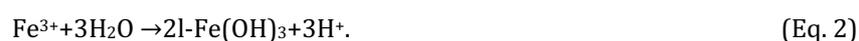

$$Fe^{3+} + 3H_2O \rightarrow 2l\text{-Fe(OH)}_3 + 3H^+. \qquad \text{(Eq. 2)}$$

The pH was maintained at 7.5 ± 0.1, by adding the last 20 mL of KOH, 500 µL at a time. Amorphous ferrihydrite-containing precipitates are prone to further phase modification, even if stored as dry powders or immersed in water at ambient temperature [17]. For this reason, the product was not oven-dried, instead, centrifuged at 10000 rpm for 15 minutes and subsequently freeze dried for 2 days. The identity of synthesised (2l-Fe(OH)$_3$(s) and γ-FeOOH(s)) and purchased (α-FeOOH(s), α-Fe$_2$O$_3$(s) and α-Fe$_3$O$_4$(s), Thermo Fisher Scientific, Waltham, MA, USA) iron (hydr)oxide reference standards were confirmed by XRD (Supporting information, Figure S2).

### 2.2. Precipitation experiments

Supersaturated iron stock solutions were prepared by pipetting 5 mL of 1 M FeCl$_3$·6H$_2$O(cr) in 2 wt.% reagent grade HNO$_3$ into 245 mL of concentrated NaOH solutions to yield a final pH of 13.0, 13.5 and 14.0. The initial concentration of dissolved iron was thus constant at 20 mM across all experiments conducted. Stock solutions were aged at ambient temperature in fresh polyethylene containers that were rinsed repeatedly with UPW (18.2 MΩcm) before use and stirred continuously throughout the duration of the experiment. The type, shape and crystallinity of iron oxides are heavily influenced by their local chemical environment and the presence of other aqueous species that may be incorporated into the phase [36, 37]. To assess the dependence of their phase assemblage on the iron source utilised, selected stock solutions were prepared from 1 M Fe(NO$_3$)$_3$· 9H$_2$O(cr) at the same degree of supersaturation.

Analogous to the post-synthesis treatment of 2-line ferrihydrite, solid precipitates were extracted from stock solutions by centrifugation and subsequent freeze drying. They were stored as dry powders and their compositions were analysed after 20 minutes to 30 days by means of XAS, TGA and XRD. Here, each point in time corresponds to a freshly prepared stock solution. Aqueous iron concentrations were investigated by ICP-OES throughout the first 2 hours of the experiment. The concentration at each point in time corresponds to the arithmetic mean of concentrations extracted from three independently prepared stock solutions. Prior to analysis, aliquots of ~1 g were taken from aged alkaline solutions, immediately filtered using 0.20 μm nylon filters (Semadeni AG, Ostermundigen, Switzerland) and acidified in 2 wt.% $HNO_3$ in UPW at a ratio of 1:10 wt.% to prevent further precipitation. The matrix was prepared from 65 wt.% $HNO_3$ (EMSURE®), from Merck Group (Merck KGaA, Darmstadt, Germany). No more than 10 aliquots of ~1 g were taken from the same reservoir.

## 2.3. Analytical methods

### 2.3.1. Scanning electron microscopy
All SEM images were obtained at the ETH Zürich Scientific Center for Optical and Electron Microscopy (ScopeM) using a Thermo Fisher Scientific Quanta 200F at high vacuum and 15 kV voltage.

### 2.3.2. Synchrotron-based investigations
XAS spectra at the Fe K-edge were collected at the PHOENIX I beamline (Photons for the Exploration of Nature by Imaging and XAFS) at the Swiss Light Source (SLS), Paul Scherrer Institute (PSI), Villigen, Switzerland. The beamline allows for measurements in the tender X-ray region, ranging from 0.8 to 8.0 keV. The photon source is a linearly and elliptically polarisable APPLE II undulator, granting a flux of $\sim 10^{12}$ photons s$^{-1}$ at 3 keV.

Measurements were conducted under vacuum ($\sim 1 \cdot 10^{-6}$ bar) and ambient temperature, employing two detection modes simultaneously, (i) the Total Electron Yield (TEY) and (ii) the Total Fluorescent Yield (TFY) at a beam size of $0.9 \times 0.9$ mm$^2$. The sample TFY was recorded by a four-element vortex detector. The incident flux, $I_0$, was measured as TEY signal taken from a Ni-coated polyester foil. The polyester foil was mounted on an electrically insulated holder, which was located in a separate chamber at $\sim 5 \cdot 10^{-8}$ bar located 1 m away from the sample. The monochromator (Si (111)) was calibrated by assigning the first inflection point of a reference iron foil to 7111.08 eV.

Iron oxide samples were uniformly applied to conductive carbon tape, placed on a copper sample holder that had previously been roughened by sand paper to ensure a noise-free total electron signal. Reference compounds and powders extracted during precipitation experiments were measured 3 - 5 times and their spectra were normalised and averaged using Athena interface of the IFFEFIT software package[38, 39]. No attempt has been made to smoothen the data beyond glitch removal. Solid spectra were corrected for self-absorption by comparing and adjusting the fluorescent signal to that of the TEY. Extended X-ray absorption fine edge structure (EXAFS) data were measured with a duration of ~ 24 minutes per scan, converted into $k^3$-weighted $\chi(\kappa)$ and subsequently Fourier transformed into R-space with the Kaiser-Bessel window function between 1.5 to 10.0 Å$^{-1}$. Time-resolved XAS spectra of iron (hydr)oxide precipitates were fitted to reference compounds by linear combination fitting (LCF) in $k$ space over 2.0 to 9.0 Å$^{-1}$, forcing the sum of all weightings to add up to 1. A maximum of 5 reference standard spectra, those of 2l-Fe(OH)$_3$(s), α-FeOOH(s), γ-FeOOH(s), α-Fe$_2$O$_3$(s) and α-Fe$_3$O$_4$(s), and all combinations in-between, were considered. The identity and purity of reference standards utilised was confirmed by XRD and TGA.

### 2.3.3. Thermogravimetric analysis
Substances were analysed in a Netzsch STA 449 F3 Jupiter® All samples were heated from 30 to 1000 °C at 10 °C min$^{-1}$ under a Nitrogen atmosphere and feature an initial mass of 30 $\pm$ 0.1 mg. The Nitrogen gas purge flow rate was 20 mL min$^{-1}$ and the alumina crucible mass was 325 $\pm$ 5.0 mg. To quantify the conversion of one iron oxide to another, their respective derivative thermogravimetry (DTG) curves were integrated via the tangential method [40].

### 2.3.4. X-ray diffraction

XRD measurements were performed on powders extracted from iron stock solutions. Diffraction spectra were obtained using a Bruker D8 Advance diffractometer with automatic beam optimisation in a coupled 2θ-θ configuration using Co Kα radiation (λ = 1.7902 Å) and a LynxEye XE-T detector. The primary beam divergent slit distance from the sample, irradiated length, and secondary beam anti-scatter slit distance to the sample were equal to 180, 15, and 270 mm, respectively. Both primary and secondary beams feature a Soller slit opening angle of 2.3°. The primary beam axial mask used is 10 mm wide located at a distance of 110 mm. Sample holders measure 25 mm in diameter and have a linear absorption coefficient of 50 cm$^{-1}$. Samples were measured between 4° and 80° in steps of 0.02° 2θ and their spectra analysed and compared to reference powder diffraction files of 2-line ferrihydrite[41], lepidocrocite (PDF entry 00-044-1415), goethite[42, 43], hematite (PDF entry 00-033-0664) and magnetite (PDF entry 00-019-0629) using the open source XRD and Rietveld refinement programme Profex[44]. Sizes of the coherently scattering crystal domain were estimated by applying the Scherrer equation[45]. It is assumed that their standard deviation was approximately equal to 10 % of the domain size [19, 46].

### 2.3.5. ICP-OES

Analyses were carried out using an Agilent 5110 ICP-OES (Aglient Technologies Inc., Santa Clara, CA, USA), equipped with an Agilent SPS 4 autosampler. The total amount of dissolved iron was correlated to the recorded intensity via an 8-point calibration line in-between the range of 0.01 to 50.00 mg L$^{-1}$ (Supporting information, Figure S8). Computed limits of detection (LOD) and quantitation (LOQ) as well as the elemental composition of all standards are reported in Tables S1 and S2.

The aqueous concentration of iron in equilibrium with the solid iron (hydr)oxide phases stabilised is in the orders of µg L$^{-1}$ and hence differs from the initial amount of iron dissolved by 3-4 orders of magnitude. Here, particular caution was exercised to determine the accuracy and precision of the analytical method and assess potential measurement interferences due to other high-concentration elements including Na and K present in the orders of g L$^{-1}$. To avoid spectral interferences, four analyte emission lines, namely Fe 234.350, 238.204, 239.563 and 259.940 nm were considered.

In line with the recommendations of Caruso et al. [47] the analyte's limit of detection (LOD) and quantitation (LOQ) were computed as three and ten times the standard deviation in the signal intensity of ten independently prepared blanks containing 280 mg L$^{-1}$ of Na, 200 mg L$^{-1}$ of K and S and 80 mg L$^{-1}$ of Ca. For the purpose of additional quality control, known concentrations of 0.1 mM and 0.3 mM of FeCl3·6H$_2$O(cr) in 2 wt.% HNO$_3$ were analysed, and their concentration recalculated for all spectral lines considered. It was found that the measured concentrations differed from the expected ones by less than 1 % at 234.350 and 259.940 nm, whilst the closeness of agreement was significantly lower for the remaining spectral lines, exceeding 1 and 4 % at 238.204 and 239.563 nm, respectively. Based on these figures of merit, it was decided to determine the concentrations based on the selected spectral line at 259.940 nm. A comparison of the progression of [Fe] with time as determined via the calibration curves of 234.350 and 259.940 nm is presented in Supporting information, Figure S9. The highest LOQ, i.e. 9.1 µg L$^{-1}$ was considered to be the global limit of quantitation. Correspondingly, the lowest concentration standard was chosen to be just above the global LOQ at 10 µg L$^{-1}$.

# 3. Results and discussion
## 3.1. Mineral phase identification

Scanning electron microscope images of powders extracted from supersaturated iron stock solution clearly show a transition towards more crystalline aggregates. As illustrated in Figure 1a, iron (hydr)oxide powders extracted after 20 minutes of equilibration time remain amorphous, resembling the sheet-like structure of ferrihydrite aggregates[48]. Within 30 days, the flaky amorphous material appears to have transformed completely into polycrystalline needles.

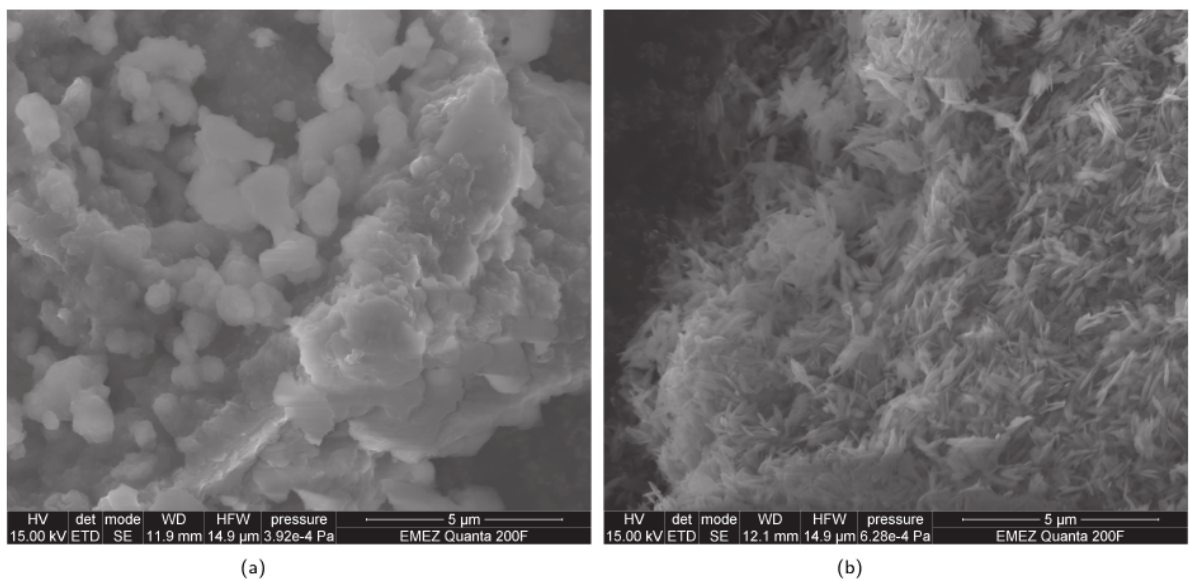

**Figure 1:** SEM images of iron (hydr)oxide powders extracted from supersaturated stock solutions at alkaline pH. Figure 1a displays amorphous aggregates extracted after 20 minutes of equilibration time, whilst Figure 1b shows needle aggregates equilibrated after 30 days.

The average oxidation state and coordination of crystalline iron bearing phases can be determined by the position and intensity of their characteristic XAS pre-edge located approx. 15 - 20 eV previous to the main K-edge[49, 50]. Whilst the average redox state largely depends on the centroid position of the pre-edge feature, integrated peak intensities, among other reasons, are determined by the degree of centrosymmetry[51]. Amongst the 6-fold oxygen coordinated $Fe^{3+}$ bearing reference components as presented in Figure 2, 2-line ferrihydrite features the highest integrated peak intensity, followed by lepidocrocite, goethite and hematite. This trend is to be expected, as peak intensities scale with the degree of octahedral $Fe(O,OH)_6$ polymerisation[49]. Fe K-edge XANES spectra of iron (hydr)oxide solids extracted from supersaturated stock solutions at pH = 14.0 (Figure 2c and Figure 2d) demonstrate a clear transition towards more centrosymmetrically coordinated $Fe^{3+}$ bearing minerals, i.e., from 2-line ferrihydrite towards a more crystalline iron hydroxide. As evident from Figure 2d, the pre-edge feature gradually reduces in intensity.

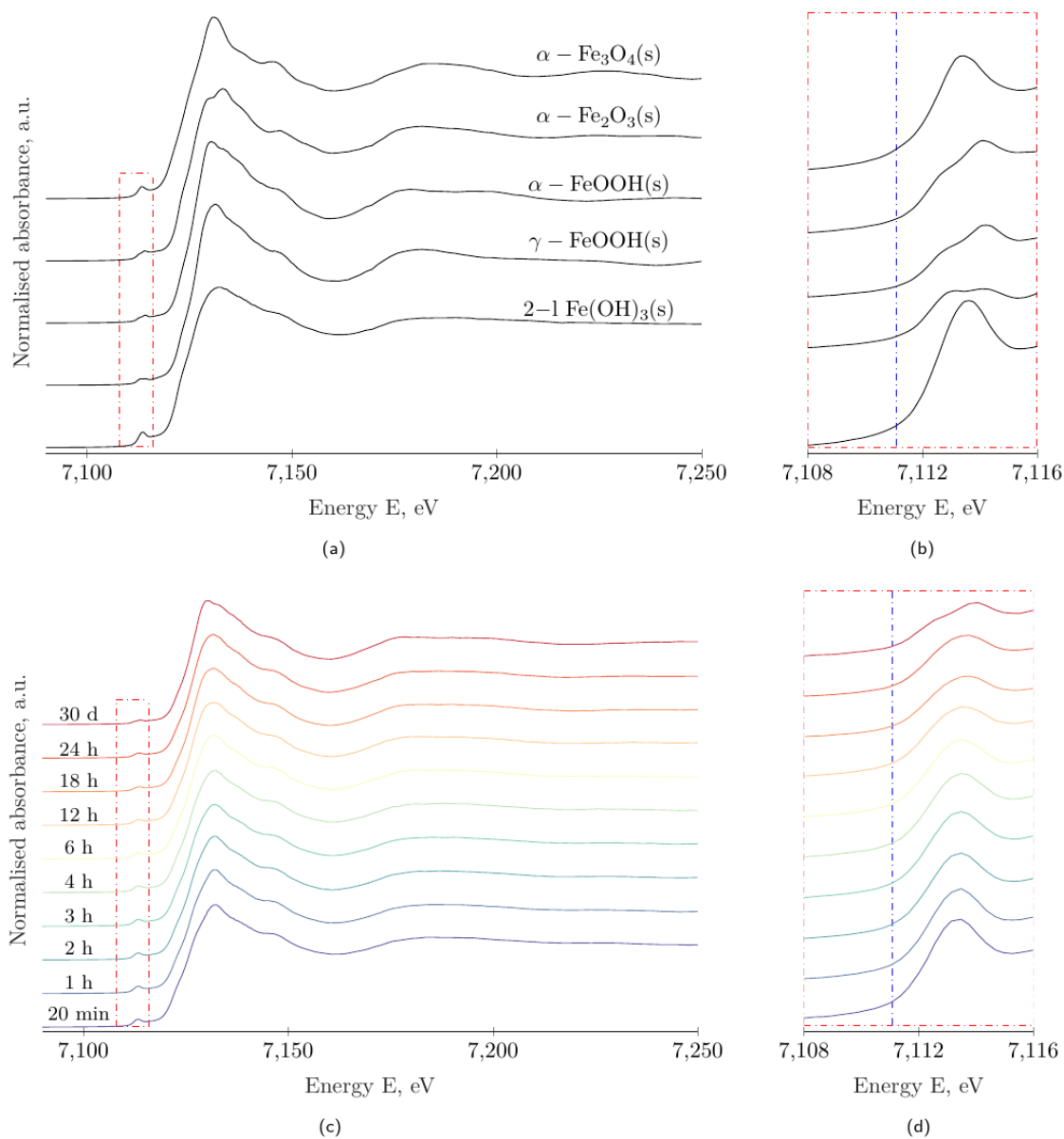

**Figure 2:** Normalised Fe K-edge XANES and pre-edge spectra of iron (hydr)oxide reference components (Figure 2a and 2b) and solids extracted from supersaturated iron stock solutions (Figure 2c and 2d) containing 20 mM $FeCl_3 \cdot 6H_2O(cr)$ at pH = 14.0. The first inflection point of a pure reference iron foil measured in the same beamline is marked as line ( $-\cdot-$ ) at 7111.08 eV.

Moreover, the relative contributions of the 1s→3d/4p [51, 52] transitions shift over time, leading first to broadening and then to splitting of the pre-edge peak. In the absence of extra transitions at ~7114 eV related to Fe clustering as visible in the reference spectrum of hematite[53], the peak splitting observed is characteristic to variants of the FeOOH(s) phase such as lepidocrocite and goethite. It is furthermore noticeable that the pre-edge centroid position does not shift as a function of equilibration time, sustaining that the redox state remained constant and no iron reduction occurred within the timespan of the experiment[49].

To quantify the extent of phase transformation, Fe K-edge XAS spectra of iron (hydr)oxide solids have been transformed into $k$-space and subsequently fitted to all combinations within the pool of reference spectra (Supporting information, Figure S1). At all equilibration times, the best fits have been universally achieved by solely including the spectra of 2-line ferrihydrite and goethite. Fe K-edge EXAFS $k^3\chi(\kappa)$ fitting results and the respective mole fraction of 2-line ferrihydrite remaining and goethite formed are displayed in Figure 3.

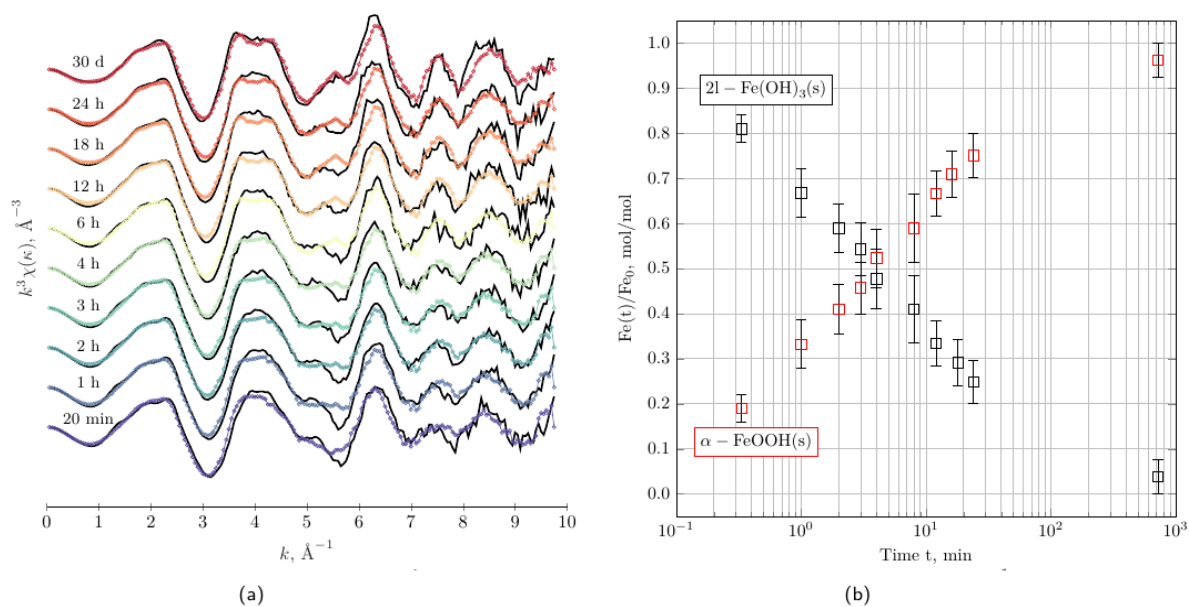

**Figure 3:** Fe K-edge EXAFS $k^3\chi(\kappa)$ (solid black lines) and corresponding linear combination fits (coloured marker lines) of aged solid samples extracted from a supersaturated $FeCl_3 \cdot 6H_2O(cr)$ stock solution at pH 14 (Figure 3a) and time-dependent fractions of the respective reference solids obtained from LCF (Figure 3b). Fits were achieved using the reference standards 2l-$Fe(OH)_3$(s) and α-FeOOH(s), and the fitting range was 2 to 9 Å$^{-1}$.

Phase changes determined via Fe K-edge EXAFS linear combination fitting are further augmented by thermogravimetric analysis and X-ray diffraction of the very same iron (hydr)oxide powders extracted. Comparison between the DTG curves of pure reference components (Figure 4), other synthetic and naturally occurring iron (hydr)oxide samples [34, 54-58] and those of the samples investigated in this study (Figure 5) suggests the stabilisation of goethite or lepidocrocite from ferrihydrite-type phases.

As shown in Figure 4, the reference 2-line ferrihydrite features a weight loss of 3.0 and 17.5 % before and exceeding 100 °C, respectively. Discrepancies between the theoretical water content of ~17 % and the total weight loss recorded can be attributed to varying extents of physisorbed water [34], arising from to differences in sample synthesis and post-treatment [34, 54, 55]. Irrespectively, the main weight loss occurs at temperatures lower than 250 °C. In contrast, the evaporation of free water from lepidocrocite and goethite occurs at temperatures < 100 °C [56]. Major weight losses of lepidocrocite due to its dehydration to hematite occur at 200 ≤ T ≤ 275 °C, respectively [56]. Goethite reference samples investigated loose the majority of their weight due to the evaporation of chemically bound water and the associated phase transformation to hematite in the same temperature interval. The exact temperature at which FeOOH(s) losses weight depends on the amount of FeOOH present in the TGA as well as its crystallite size [40].

The DTG curves of iron (hydr)oxides extracted at different equilibration times feature two distinct peaks, (i) at T ~100 °C and (ii) 200 ≤ T ≤ 300 °C, matching the temperature intervals across which (i) 2-line ferrihydrite and (ii) lepidocrocite and goethite loose the majority of their weight. As illustrated in Figure 5, the weight loss across the first region of interest (ROI), i.e., from 50 to 190 °C, relative to the area underneath both peaks decreases exponentially with time. Conversely, the fraction of weight loss occurring across the temperature interval from 200 to 300 °C increases by the same amount, as shown in Figure 5b.

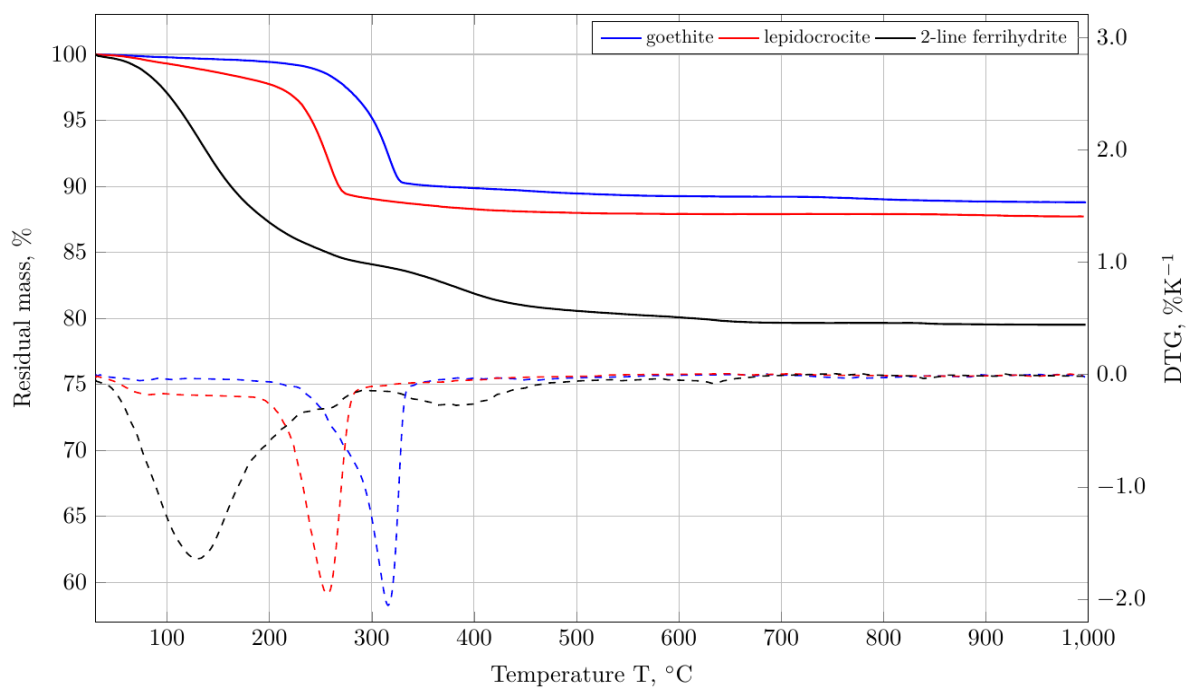

**Figure 4:** Percentage residual mass (continuous) and DTG (dashed) curves of reference components 2-line ferrihydrite, lepidocrocite and goethite.

Assuming that water losses within the first ROI (50 – 190 °C) are solely due to the decomposition of 2-line ferrihydrite, the amount of 2-line Fe(OH)$_3$(s) may be computed according to

$$2\text{l-Fe(OH)}_3 = \text{WL} \times \frac{1}{3/2 \cdot \text{MW}_{\text{H}_2\text{O}}} \qquad \text{(Eq. 3)}$$

where WL is the observed weight loss as computed via the tangential method[40] and $\text{MW}_{\text{H}_2\text{O}}$ is the molecular weight of water in g mol$^{-1}$.

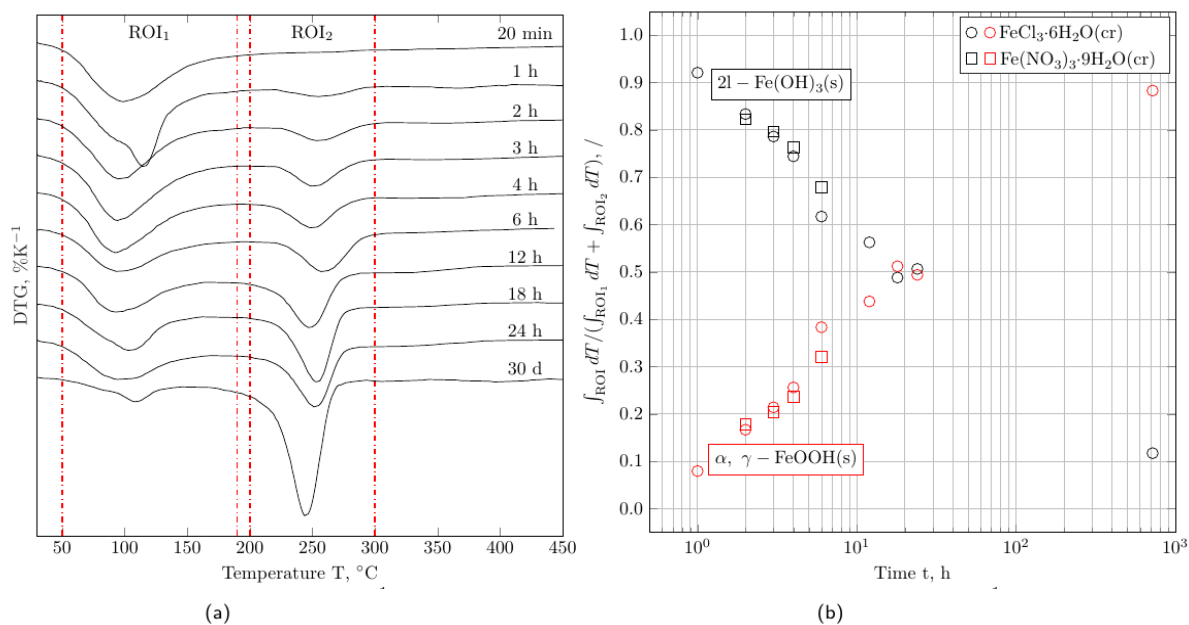

**Figure 5:** Progression of DTG curves of iron oxide samples extracted from 1 M NaOH, pH = 14.0 at different equilibration times. Solids formed from 20 mM FeCl$_3$·6H$_2$O(cr) are marked with a circle whilst those stabilised from 20 mM Fe(NO$_3$)$_3$·9H$_2$O(cr) are denoted with square markers. Figure 5a shows the DTG curves of all FeCl$_3$·6H$_2$O(cr) samples together with the ROI for both characteristic peaks observed. Figure 5b shows the area underneath each ROI, relative to the total weight loss over both peaks. All areas have been evaluated via the tangential method [40], integrated over the temperature interval [50, 190] and [200, 300], respectively.

The dehydration of α-FeOOH (goethite) and γ-FeOOH (lepidcrocite) to ½ α-Fe$_2$O$_3$ (hematite) in the second ROI (200 – 300 °C), can be quantified according to

$$\alpha\text{-}, \gamma\text{-FeOOH} = WL \times \frac{1}{1/2 \cdot MW_{H_2O}} \quad \text{(Eq. 4)}$$

As evident from Figure 6, mass fractions computed directly from the weight losses of water on the respective temperature intervals differ by about 10 % compared to mole fractions inferred by Equations 3 and 4. It is further evident that the conversion proceeds significantly more rapidly at pH = 14.0. Moreover, the reaction coordinate appears to be independent of the iron source used, as 2l-Fe(OH)$_3$(s) stabilised from both FeCl$_3$·6H$_2$O(cr) and Fe(NO$_3$)$_3$·9H$_2$O(cr) converts to FeOOH(s) at the same rate.

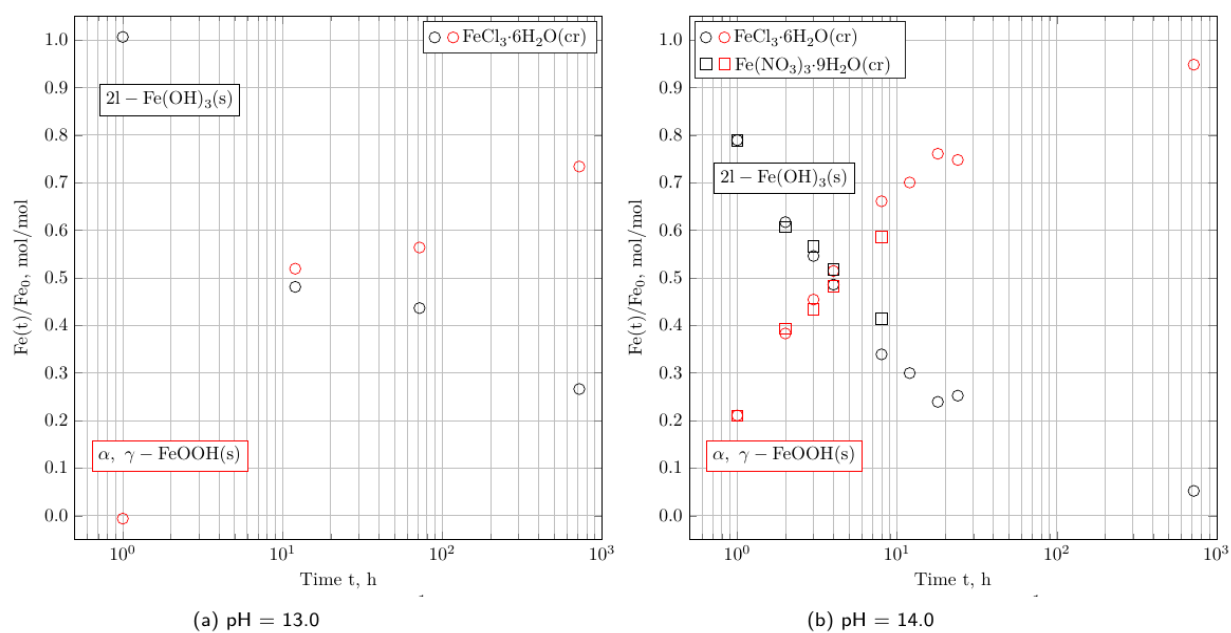

**Figure 6:** Progression of the mass fraction of Fe(OH)$_3$(s) and FeOOH(s) type iron hydroxides stabilised at pH = 13.0 (Figure 6a) and pH = 14.0 (Figure 6b) over time, as determined by TGA and computed by Equation 4. Solids formed from FeCl$_3$ · 6H$_2$O(cr) are marked with a circle whilst those stabilised from Fe(NO$_3$)$_3$·9H$_2$O(cr) are denoted with square markers.

An unambiguous characterisation of the FeOOH(s) phase is not possible via TGA, as the second ill-defined DTG peak is located within the overlapping region of both higher stability iron hydroxides.

X-ray diffractograms further demonstrate that the FeOOH(s) type iron hydroxide stabilised from 2-line ferrihydrite is goethite (α-FeOOH(s)). As displayed in Figure 7, the solids extracted after 20 minutes from supersaturated iron stock solutions feature two broad peaks centred at 2θ of ∼ 40° and ∼ 74°, similar to those of the reference powder diffraction file [41] and the synthesised 2-line ferrihydrite sample (Supporting information, Figure S3). Over time, the characteristic 110 and 111 peaks of goethite at a 2θ of ∼ 24° and ∼ 42° (Supporting information, Figure S5) emerge from the amorphous diffractograms initially recorded. It is also evident that both 2 line-ferrihydrite and goethite are formed within the first few hours of equilibration. Subsequently, the crystallinity of the final product goethite increases and no other (iron-bearing) phase is formed. This trend holds true, irrespective of the iron source used during precipitation experiments. Both, the time series for batches using FeCl$_3$·6H$_2$O(cr) and Fe(NO$_3$)$_3$·9H$_2$O(cr) feature peaks of their respective residual crystalline side product halite and sodium nitrate, as marked by the letters $h$ and $n$.

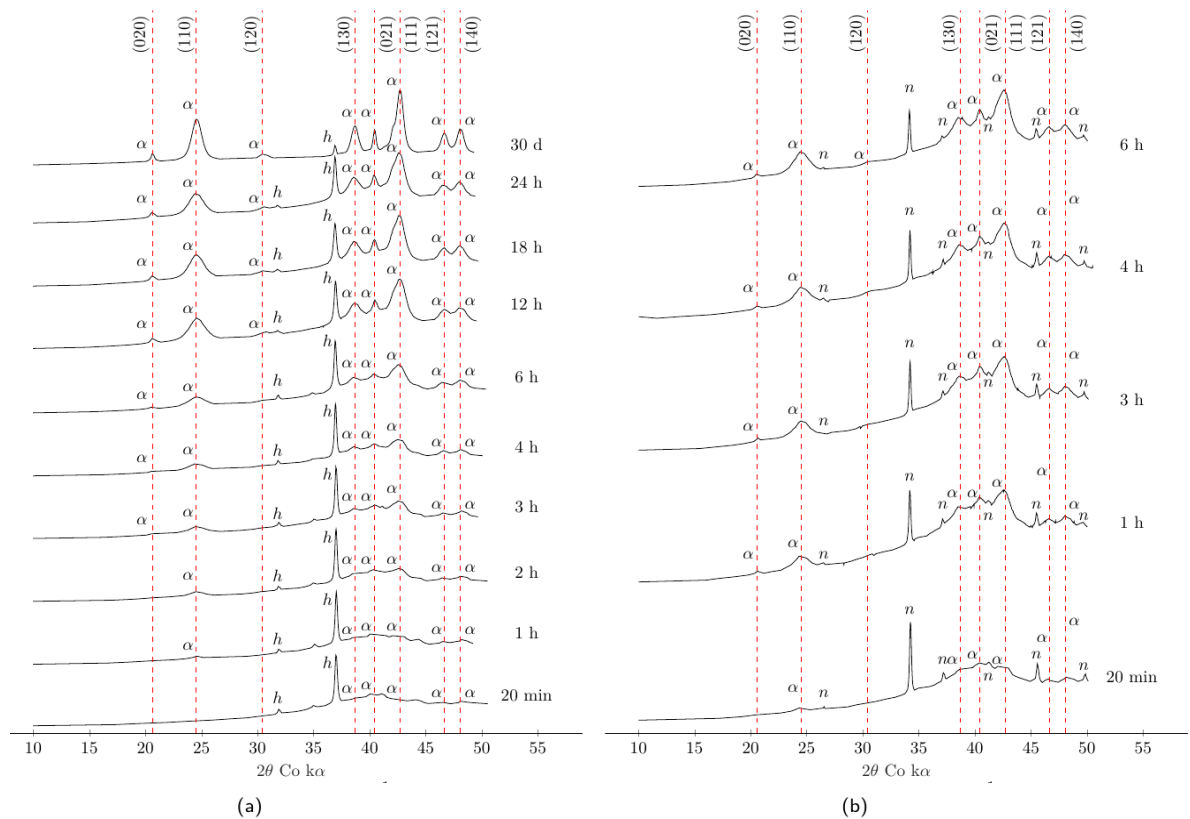

**Figure 7:** XRD patterns of time dependent iron (hydr)oxide phases formed from 20 mM FeCl$_3$·6H$_2$O(cr) (Figure 7a) and 20 mM Fe(NO$_3$)$_3$·9H$_2$O(cr) (Figure 7b) in 1 M NaOH (pH 14). Position of the main peaks of goethite (α-FeOOH(s)) are marked as — · —, whilst the peaks of halite (NaCl) and sodium nitrate (NaNO$_3$) are denoted by $h$ and $n$.

The incident transformation of the crystal structure to the orthorhombic lattice of goethite is concluded within 30 days. XRD peak analysis further confirms that the size of the coherently scattering crystal domain of goethite grows continuously over the timespan investigated. Figure 8 illustrates the increase in domain size along the 110 and 140 directions as a function of ageing time.

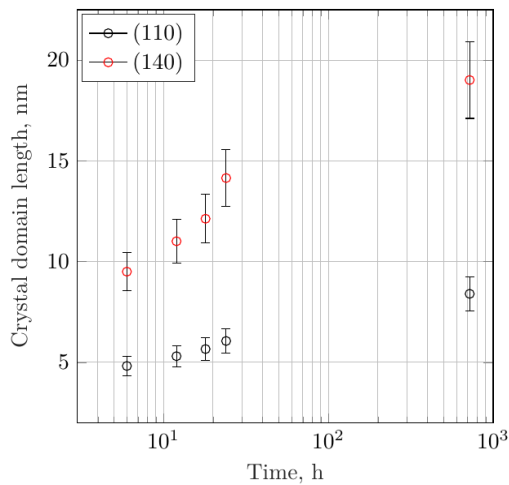

**Figure 8:** Size of the coherently scattering crystal domain of selected goethite (α-FeOOH(s)) peaks at pH 14, as estimated via peak-shape analysis and the application of the Scherrer equation[45].

### 3.2. Quantification of the total aqueous iron concentration

The total iron concentration as a function of time and pH as determined by ICP-OES is displayed in Figure 9. Irrespective of the pH, iron concentrations decrease from 20 mM to the orders of tens of µM within the first minute. Across this timespan, initial precipitation rates are estimated to be $(3.331 \pm 0.004) \cdot 10^{-4}$ mol l$^{-1}$ s$^{-1}$, irrespective of the pH. After this rapid initial decrease, the progression of [Fe] flattens out, striving towards some finite, pH dependent solubility limit. Whilst iron concentrations at pH = 14.0 remain in the orders of $10^{-5}$ M, those at pH = 13.0 are below the limit of quantification after 30 minutes. This is expected as the concentration of Fe(III) decrease and the degree of supersaturation with respect to any of the iron (hydr)oxides increases with decreasing pH.

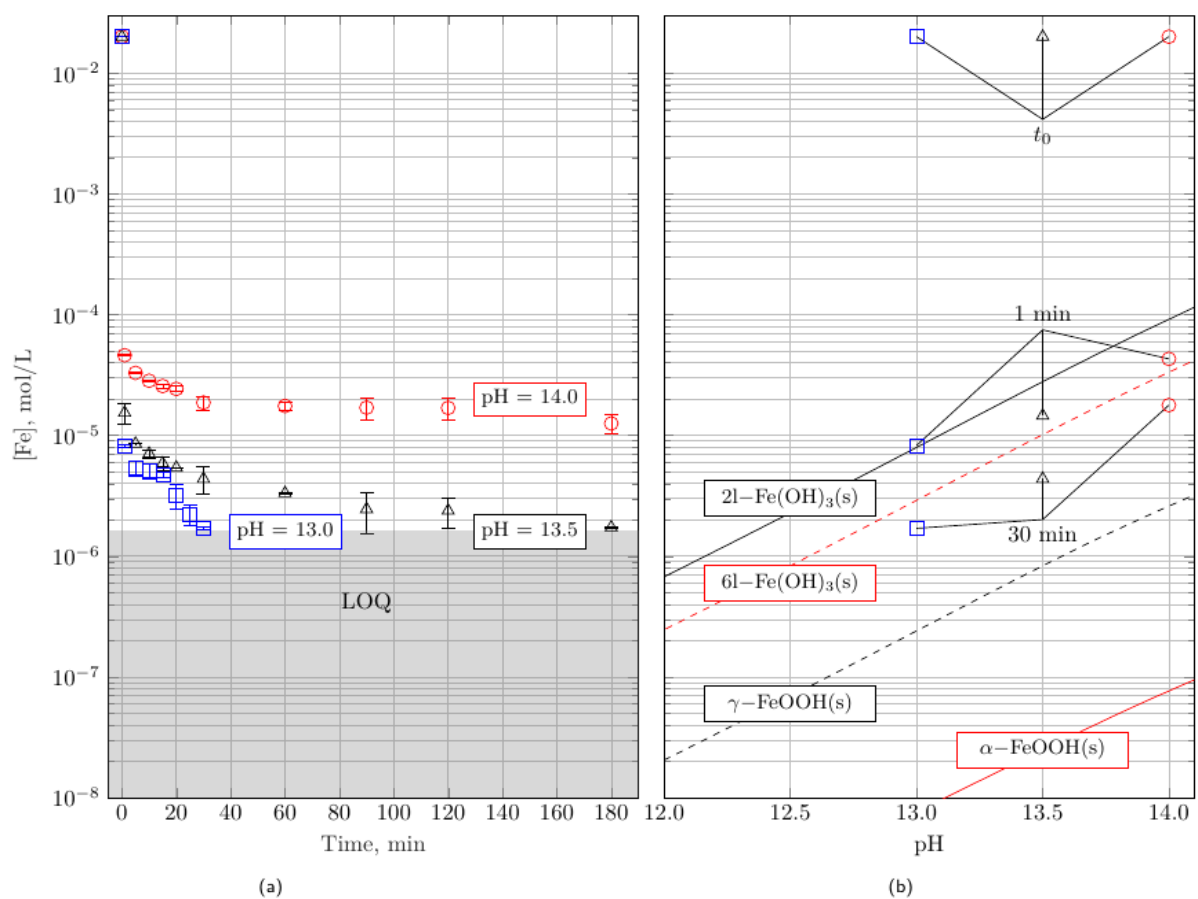

**Figure 9:** Total iron concentration as measured by ICP-OES. Figure 9a displays the progression of [Fe] in mol/L over time for pH 13.0, 13.5 and 14.0 at ambient temperature. Error bars represent the standard deviation in the measurements of three independently prepared solutions. Figure 9b shows the concentration at selected points in time versus the solubility of Fe(III), as controlled by the solubility of 2- and 6- line ferrihydrite (2-, 6-l Fe(OH)$_3$(s)), lepidocrocite (γ-FeOOH(s)) and goethite (α- FeOOH(s)) [32].

Concentration measurements are generally consistent across all precipitation experiments and spectral lines at each time step. After 1 minute and longer, the measured iron concentrations were at or below the solubility of 2-line ferrihydrite. After 30 minutes all concentrations were below the solubility of 6-line ferrihydrite, which was not observed to form by either XAS or XRD. As time progresses, concentrations align parallel to the solubility limits of various iron (hydr)oxides and the system remains supersaturated with respect to lepidocrocite, goethite and other high-stability phases. This implies that at least one of the iron (hydr)oxides formed features a solubility lower than that of 6-line ferrihydrite. The time-resolved experimental techniques employed here demonstrate that the formation of stable iron (hydr)oxide phases at alkaline pH comprises the initial rapid precipitation of amorphous 2-line ferrihydrite as intermediate phase followed by its dissolution and the crystallisation of goethite from solution. The formation of 6-line ferrihydrite and lepidocrocite, which were both oversaturated during the experiments, was not observed indicating that their formation kinetics are slower at pH 13 to 14.

### 3.3. Kinetics of 2-line ferrihydrite transformation

To compare the transformation rates measured here at highly alkaline pH to those measured in acidic and circumneutral environments, time-dependent concentration profiles of 2-line ferrihydrite obtained by EXAFS linear combination fitting and those measured by Schwertmann et. al. [20], Schwertmann and Murad [21] and Das et. al. [26] were plotted together. The rate constants were calculated by the integrated first order rate equation

$$\text{Fe}(t) = \text{Fe}_0 \cdot e^{-kt}. \tag{Eq. 5}$$

Figure 10 shows the calculated first order rate constant k in h$^{-1}$ as a function of the pH. It can be recognised that the rate of transformation of 2-line ferrihydrite to goethite (and hematite at 5 ≤ pH ≤ 10) strictly increases as a function of the activity of OH$^-$ or pH. In acidic to neutral conditions, the rate constants derived based on the data of Schwertmann and co-workers[20], [21] vary by less than one order of magnitude, reaching values of $8.1 \cdot 10^{-5}$ h$^{-1}$ at pH = 2 and $2.5 \cdot 10^{-4}$ h$^{-1}$ at pH = 7, respectively.

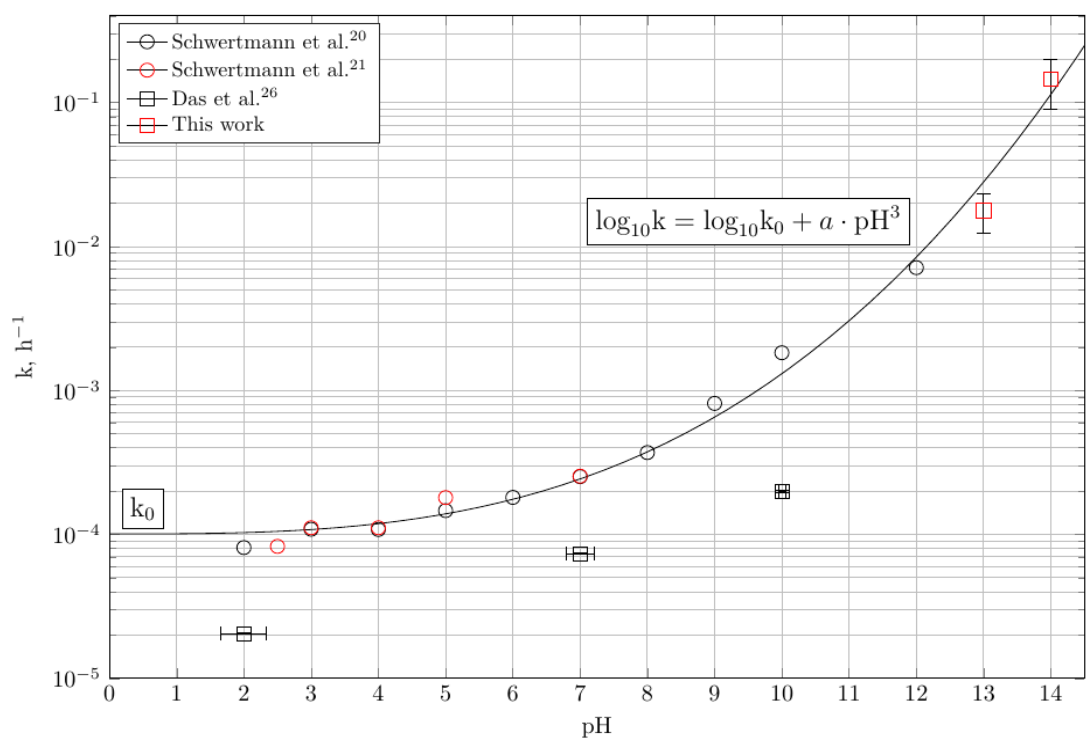

**Figure 10:** Estimated first order rate constants of 2-line ferrihydrite transformation calculated by Das et. al. [26] and obtained by fitting Equation 5 to the progression of the molar fraction of 2-line ferrihydrite at 23±2 °C, as determined by EXAFS LCF, TGA and measured by Schwertmann et. al. [20] and Schwertmann and Murad [21]. The rate constant obtained in this work at pH = 14.0 represents the average rate constant and error achieved by fitting the estimated mol fractions from EXAFS LFC and TGA, while the rate constant at pH = 13.0 is derived from TGA only, assuming the same uncertainty. Plots of the fits achieved are shown in Figure S10 and S11.

Various k calculated by Das et. al. [26] are slower than those calculated by re-evaluating the data of Schwertmann et. al. [20] by approximately one order of magnitude due to unknown reasons. Nevertheless, they follow the same pH dependence. At pH = 14, the here estimated first order rate constant surpasses that at pH = 2 to 4 by more than 3 orders of magnitude, reaching a value of $1.5 \cdot 10^{-1}$ h$^{-1}$. This drastic increase in phase transformation rates across the pH range investigated is well captured by

$$\log_{10} k = \log_{10} k_0 + a \cdot pH^3. \qquad (Eq.\ 6)$$

where $k_0 = 1.0076 \cdot 10^{-4}$ h$^{-1}$ is the standard rate constant at pH = 0 and $a = 0.0011$ is an empirical parameter at 25°C. It must be emphasised that the analysis presented in this section does not permit to draw any conclusion regarding the rate of 2-line ferrihydrite formation or the growth mechanism of goethite. Instead, we provide a semi-empirical relationship describing the overall transformation rate as a function of the pH, and therefore, the OH$^-$ activity. As evident from the SEM images presented in Figure 1, the mechanism leading to the formation of goethite involves changes in particle morphology and surface area. Moreover, goethite particle growth is known to be inhibited in the presence of both Si and Cl [1]. To account for the effects of other physiochemical parameters in addition to the solution pH and formulate a crystallisation mechanism that describes the growth of both minerals in partial equilibrium with the aqueous phase requires careful construction of all kinetic rate laws involved. A formulation of such mechanism will be the subject of future work.

### 3.4. Environmental implications

Observations demonstrate that, despite fast dissolution kinetics of 2-line ferrihydrite at high pH values, the aqueous concentration of Fe(III) decreases only slowly and the solutions remain supersaturated with respect to goethite for a significant time. Due to its very rapid formation and slower dissolution, 2-line ferrihydrite can be considered a point source of Fe(III) that maintains the aqueous phase in a state of disequilibrium. Within the alkaline pore solution of cementitious matrices such as those used in radioactive waste storage, the amount of Fe(III) above the solubility limit of goethite can readily be transported across the pore network or taken up by any other cementitious phase in the system, prospectively impacting their service life and long-term ability to contain hazardous radionuclides[33, 59]. As the transformation from amorphous 2-line ferrihydrite to crystalline goethite coincides with a tenfold reduction in the specific surface area, also the capacity to immobilise toxic elements such as As, Sr or Cd is expected to be severely reduced. The estimated 2-line ferrihydrite half-life $t_{1/2}$ at pH = 10, i.e. at mildly alkaline conditions characteristic to uranium mine tailings [4, 26], amounts to approximately 15 days at 25 °C. At pH > 13, i.e. the pH characteristic to radioactive waste tailings in Portland cements, ferrihydrite half-life is approximately 40 hours and 5 hours at pH 14, as present in alkali activated cements. However, the rate of 2-line ferrihydrite transformation also depends on the presence of other multivalent impurities that may impede sorption of primary elements of concern [2, 17, 34, 35]. A rigorous comparison between these competing phenomena requires a more thorough understanding of both the crystallisation process including the initial precipitation of 2-line ferrihydrite, the growth of goethite and the change in particle morphology.

### Acknowledgements

All XAS measurements were performed at the PHOENIX beamline at the Swiss Light Source, Paul Scherrer Institut, Villigen, Switzerland. The authors would like to thank Dr. Michael Plötze, Annette Röthlisberger and Marion Rothaupt from the ClayLab, Institute for Geotechnical Engineering, ETH Zürich for access and help with the XRD sample preparation, data evaluation and valuable discussions. The authors would also like to thank Dr. Alexander German for TGA measurement as well as Dr. Yiru Yan and Raphael Kuhn from the Empa Concrete & Asphalt Laboratory for their immense support in preparing and conducting the ICP measurements. The authors are grateful to the European Research Council (ERC) for the financial support provided under the European Union's Horizon 2020 research and innovation program (grant agreement no. 848794). The support from ETH Zurich internal Funding and NSF CMMI 1728358 helped enable the collaboration between ETH Zurich and Oregon State University.

**Author contributions**

Fabio E. Furcas, Shishir Mundra, Ueli M. Angst, and Barbara Lothenbach conceived the overall study; all authors contributed to the study design. Experimental work was performed by Fabio E. Furcas, Camelia Borca, Shishir Mundra, Cristhiana C. Albert and Thomas Huthwelker. Fabio E. Furcas, Barbara Lothenbach, Shishir Mundra, Camelia Borca, Ueli M. Angst and O. Burkan Isgor, performed the analysis and interpretation of the results. Fabio E. Furcas wrote the main draft of the manuscript, to which all authors contributed. Ueli M. Angst was the main supervisor of the project. All authors read and approved the final manuscript.

**Supporting information**

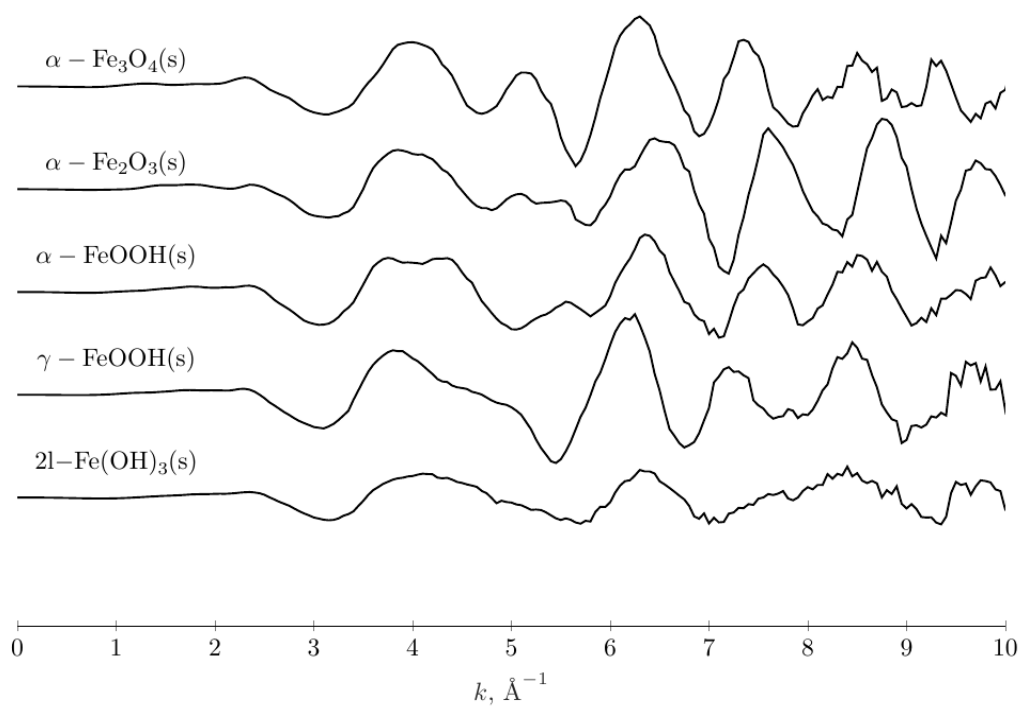

**Figure S1:** Fe K-edge EXAFS $k^3\chi(\kappa)$ spectra of pure iron (hydr)oxide reference standards 2-line ferrihydrite (2l-Fe(OH)$_3$(s)), lepidocrocite (γ-FeOOH(s)), goethite (α-FeOOH(s)), hematite (α-Fe$_2$O$_3$(s)) and magnetite (α-Fe$_3$O$_4$(s)).

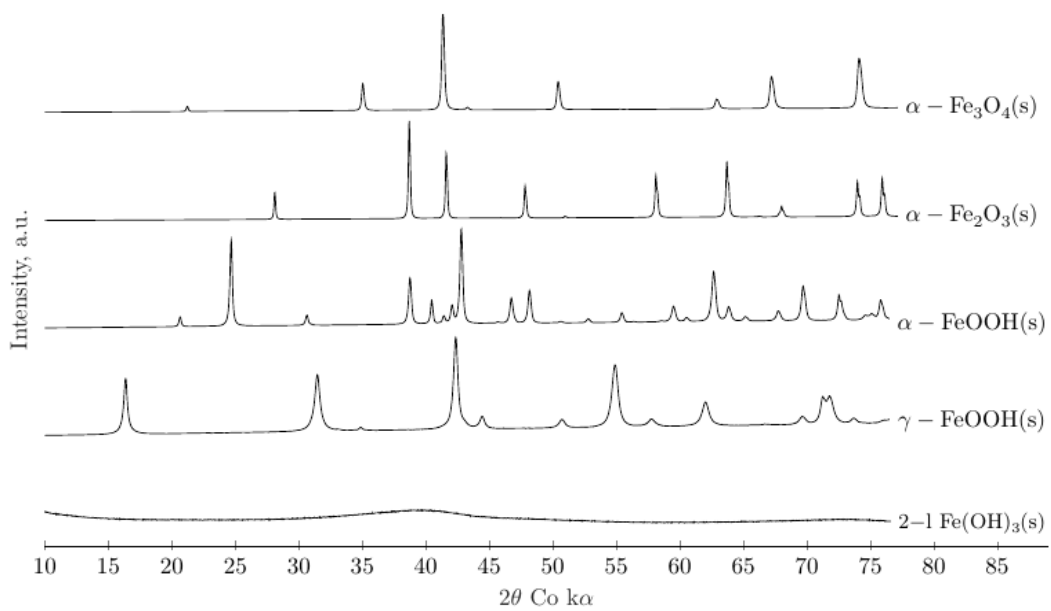

**Figure S2:** X-ray diffractograms of pure iron (hydr)oxide reference standards 2-line ferrihydrite (2l-Fe(OH)$_3$(s)), lepidocrocite (γ-FeOOH(s)), goethite (α-FeOOH(s)), hematite (α-Fe$_2$O$_3$(s)) and magnetite (α-Fe$_3$O$_4$(s)).

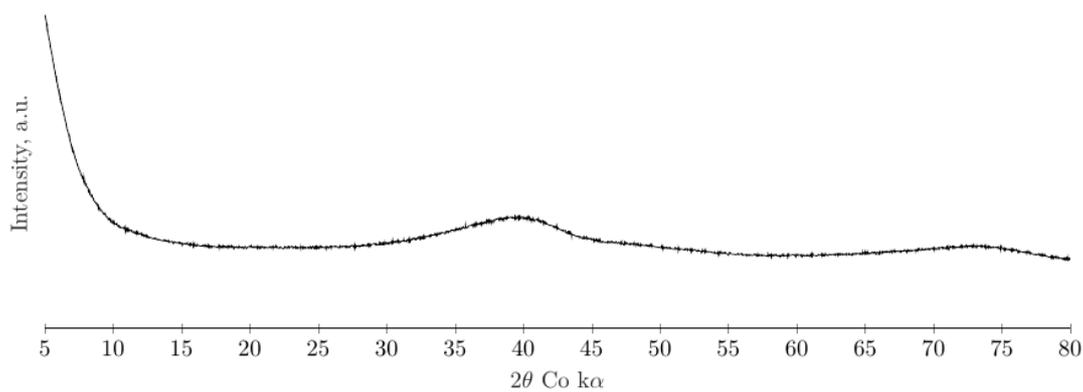

**Figure S3:** X-ray diffractogram of 2-line ferrihydrite (2l-Fe(OH)$_3$(s)).

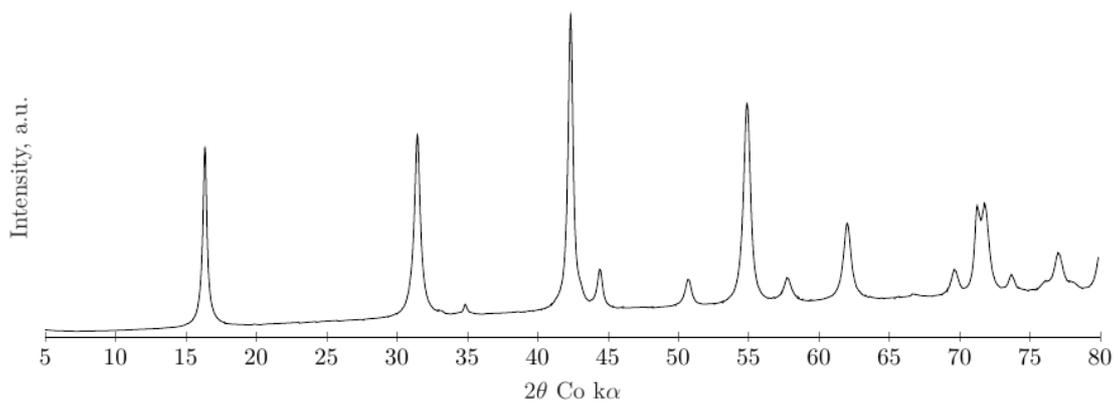

**Figure S4:** X-ray diffractogram of lepidocrocite (γ-FeOOH(s)).

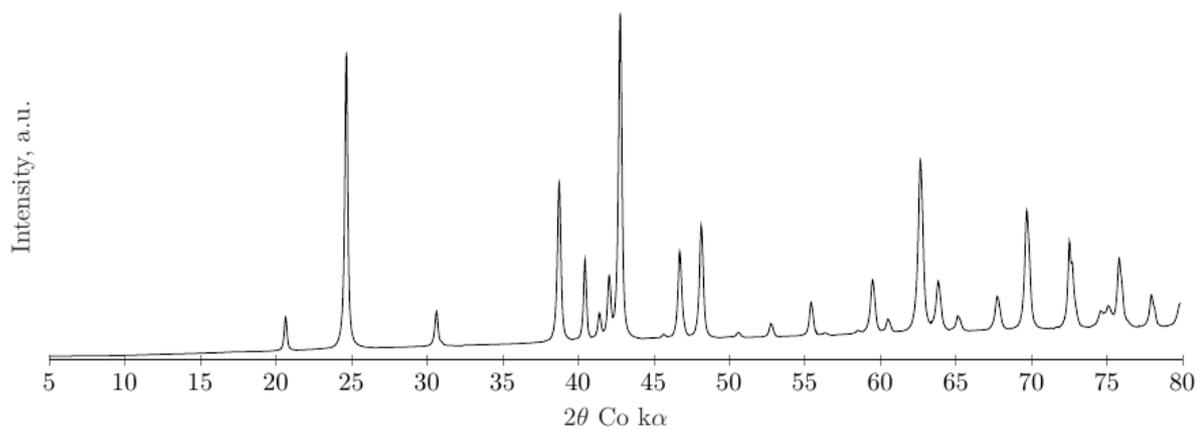

**Figure S5:** X-ray diffractogram of goethite (α-FeOOH(s)).

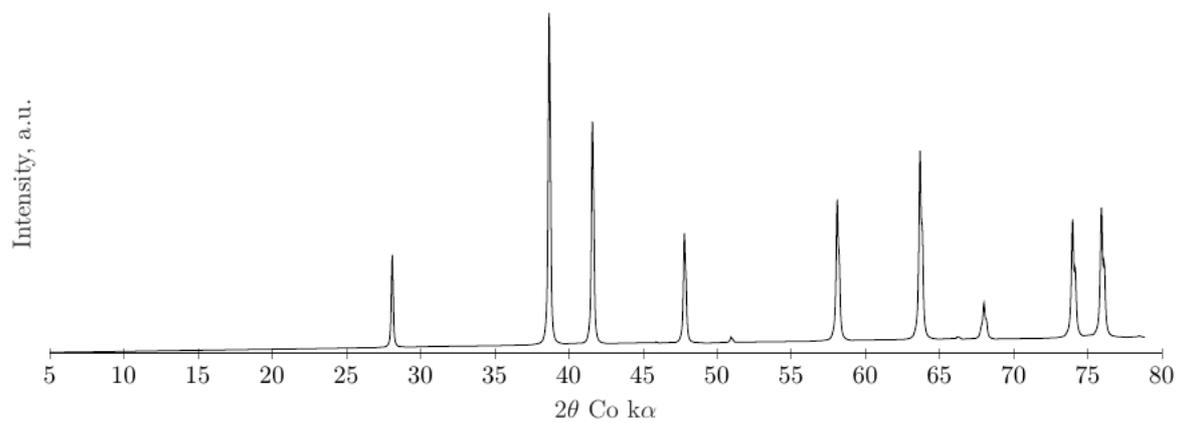

**Figure S6:** X-ray diffractogram of hematite (α-Fe$_2$O$_3$(s)).

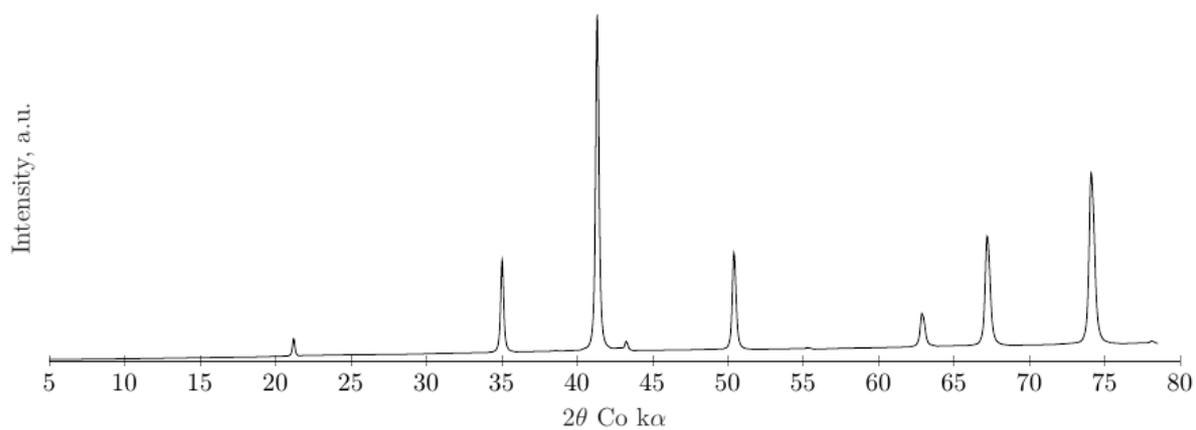

**Figure S7:** X-ray diffractogram of magnetite (α-Fe$_3$O$_4$(s)).

**Table S1:** LOD and LOQ in µg L⁻¹ and µmol L⁻¹ for the low-concentration element Fe, as determined by the recommendations of Caruso et al. [47]. Concentrations are obtained by means of a linear 8-point interpolation containing 0.01, 0.10, 0.50, 1.00, 5.00, 10.00, 25.00 and 50.00 ppm of the measured elements.

| Element | Spectral line, nm | SD | LOD | | LOQ | |
|---|---|---|---|---|---|---|
| | | (ppm) | (µg L⁻¹) | (µmol L⁻¹) | (µg L⁻¹) | (µmol L⁻¹) |
| Fe | 234.350 | $2.961 \cdot 10^{-4}$ | 0.888 | 0.016 | 2.961 | 0.053 |
| | 238.204 | $4.726 \cdot 10^{-4}$ | 1.418 | 0.025 | 4.726 | 0.085 |
| | 239.563 | $4.348 \cdot 10^{-4}$ | 1.304 | 0.023 | 4.348 | 0.078 |
| | 259.940 | $9.081 \cdot 10^{-4}$ | 2.724 | 0.049 | 9.081 | 0.163 |

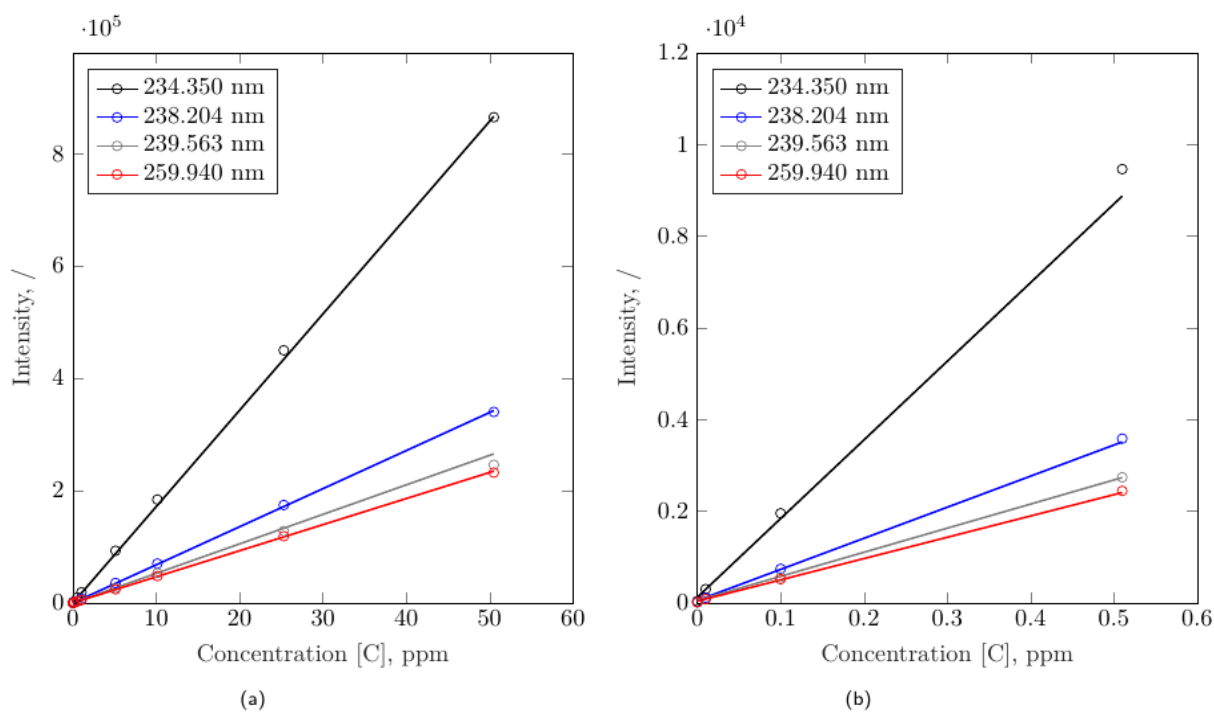

**Figure S8:** ICP-OES calibration curves for all measured Fe spectral lines employing a linear 8-point fit at 0.01, 0.10, 0.50, 1.00, 5.00, 10.00, 25.00 and 50.00 ppm. Fits of all spectral lines feature a coefficient of determination $R^2 \geq 0.999$. Figure S8a displays the full calibration range, whilst Figure S8b shows a zoomed-in section of the three lowest standards employed above the blank. Note that the highest estimated limit of quantitation (LOQ) of ~0.91 ppm at 259.940 nm as listed in Table S1 lies just below the lowest standard employed.

**Table S2:** Elemental composition of the ICP calibration solutions used to determine the total amount dissolved aqueous iron. For all runs, the blank solution consisted of 2 wt.% $HNO_3$ in UPW.

| Standard Number | Element, ppm | | | | | | | | |
|---|---|---|---|---|---|---|---|---|---|
| | Na | K | Ca | Mg | Al | Fe | S | Si | P |
| 8 | 50.5506 | 50.2203 | 50.3531 | 50.4780 | 20.2052 | 50.3020 | 50.2344 | 50.3595 | 50.3269 |
| 7 | 25.3110 | 25.1456 | 25.2121 | 25.2747 | 10.1169 | 25.1865 | 25.1527 | 25.2153 | 25.1990 |
| 6 | 10.1346 | 10.0684 | 10.0950 | 10.1201 | 4.0508 | 10.0848 | 10.0712 | 10.0963 | 10.0898 |
| 5 | 5.0766 | 5.0435 | 5.0568 | 5.0693 | 2.0291 | 5.0517 | 5.0449 | 5.0574 | 5.0542 |
| 4 | 1.0239 | 1.0172 | 1.0199 | 1.0224 | 0.4093 | 1.0188 | 1.0175 | 1.0200 | 1.0194 |
| 3 | 0.5096 | 0.5063 | 0.5076 | 0.5089 | 0.2037 | 0.5071 | 0.5064 | 0.5077 | 0.5640 |
| 2 | 0.1017 | 0.1010 | 0.1013 | 0.1016 | 0.0407 | 0.1012 | 0.1011 | 0.1013 | 0.1013 |
| 1 | 0.0101 | 0.0101 | 0.0101 | 0.0102 | 0.0041 | 0.0101 | 0.0101 | 0.0101 | 0.0101 |

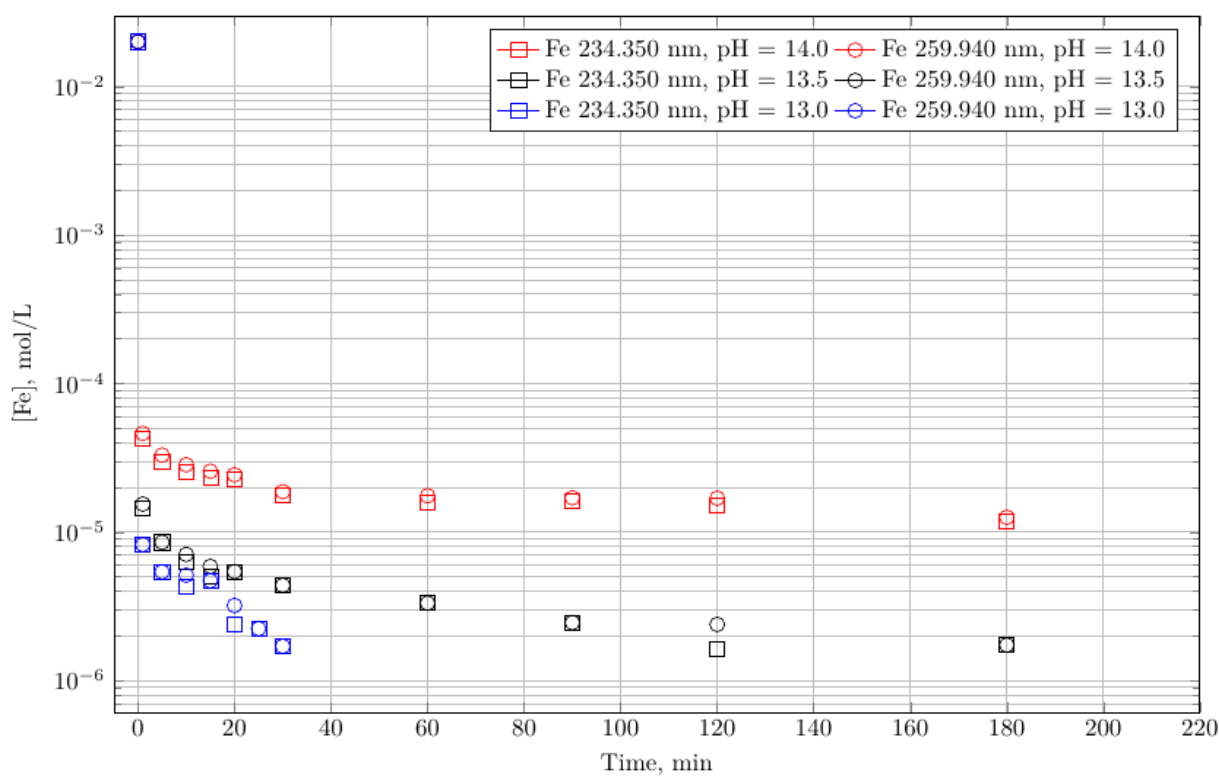

**Figure S9:** Comparison of the total amount of dissolved iron over time at various pH, as determined via calibration at the emission lines of 235.350 and 259.940 nm. Note that the concentrations as predicted via interpolation at one spectral line are within the standard deviation of three independent measurements of the other one.

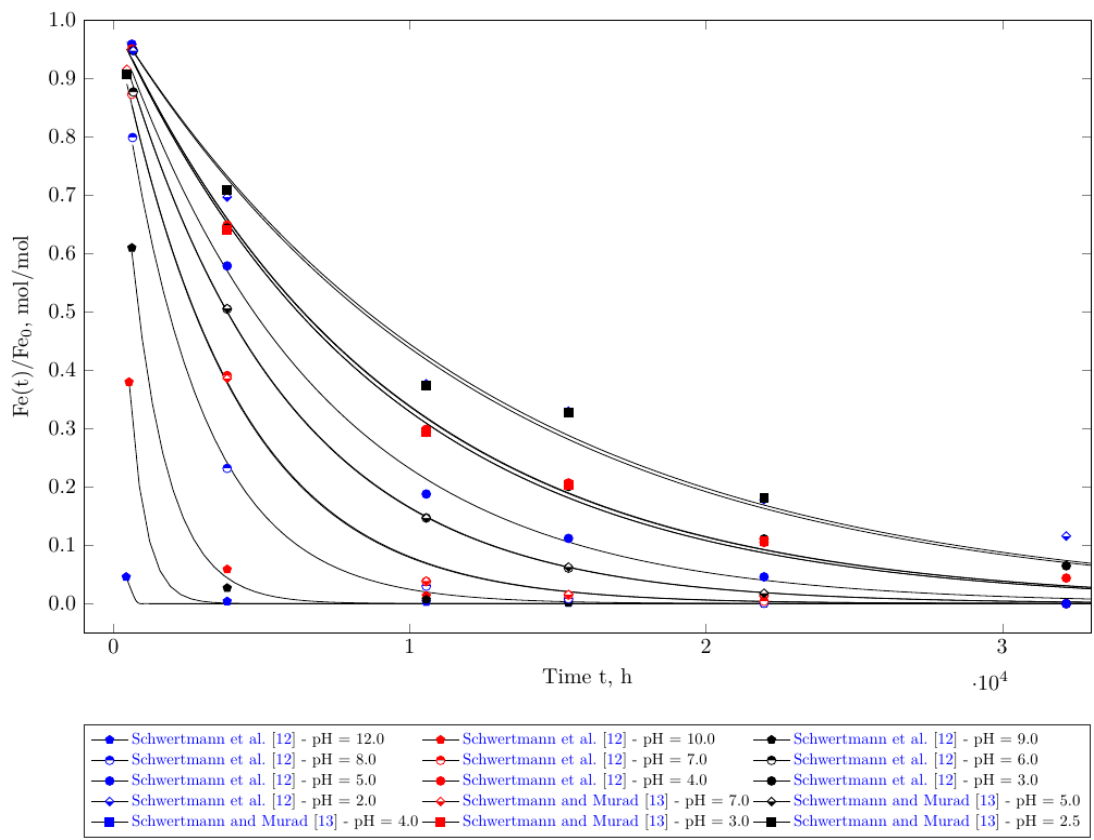

**Figure S10:** Ratios of 2-line ferrihydrite Fe(t) relative to the initial iron concentration $Fe_0$ over time at various pH.

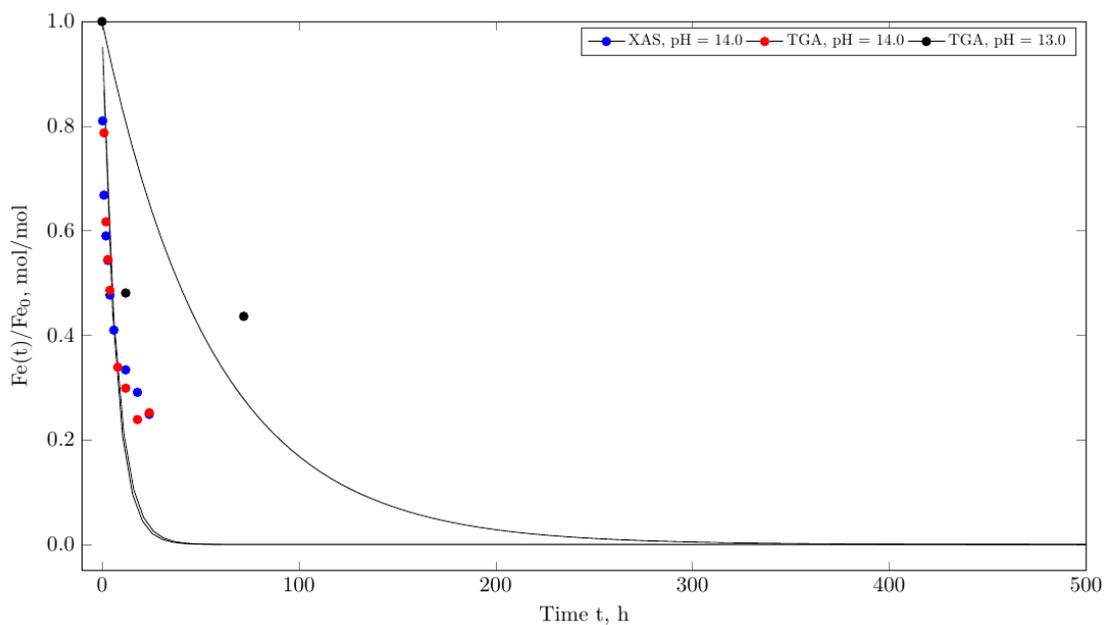

**Figure S11:** Ratios of 2-line ferrihydrite Fe(t) relative to the initial iron concentration $Fe_0$ over time, as determined by XAS LCF at pH = 14.0 and TGA at pH = 13.0 and 14.0.